\documentclass[preprint,12pt]{elsarticle}

\usepackage{amssymb,algorithm,algorithmic,graphicx,color}

\usepackage{epsfig}
\usepackage{amssymb}

\newcommand{\ra}{\rightarrow}

\newcommand{\scst}{\scriptscriptstyle}
\newcommand{\ti}{\times}

\newtheorem{definition}{Definition}[section]
\newtheorem{lem}{Lemma}[section]
\newtheorem{thm}{Theorem}[section]

\newtheorem{exmp}[lem]{Example  }

\newtheorem{alg}[lem]{Algorithm}

\begin{document}
\begin{frontmatter}
\title{Chain Homotopies for Object Topological Representations}
\author{R. Gonz\'{a}lez--D\'{\i}az}
\ead{rogodi@us.es}
\ead[url]{http://www.personal.us.es/rogodi}
\author{M.J. Jim\'{e}nez}
\ead{majiro@us.es}
\ead[url]{http://ma1.eii.us.es/miembros/majiro}
\author{B. Medrano}
\ead{belenmg@us.es}
\ead[url]{http://ma1.eii.us.es/miembros/medrano}
\author{P. Real}
\ead{real@us.es}
\ead[url]{http://www.pdipas.us.es/r/real}
\address{Depto. de Matem\'{a}tica Aplicada I, Escuela Superior de
Ingenier\'{\i}a Inform\'{a}tica, Universidad
de Sevilla, Avda. Reina Mercedes, s/n, 41012, Sevilla (Spain)}

\begin{abstract}
This paper presents a set of  tools to compute topological information of simplicial complexes, tools that are
applicable to extract topological information from digital pictures.
A simplicial complex is encoded
in a (non-unique) algebraic-topological format called AM-model. An AM-model for a given object $K$ is determined
by a concrete chain homotopy and it
provides, in particular,  integer (co)homology generators of $K$ and
representative (co)cycles of these generators.
An algorithm for computing an AM-model and the cohomological invariant $HB1$ (derived from the rank of the cohomology ring)
with integer coefficients for a finite simplicial complex
in any dimension is designed here,
extending the work done in \cite{GR05} in which the ground ring was a field.
A concept of generators which are ``nicely" representative is also presented.
Moreover, we extend the definition of AM-models to 3D binary digital images and
  we
 design  algorithms to update the AM-model information after voxel
 set operations (union, intersection, difference and inverse).
 \end{abstract}


\end{frontmatter}

\section{Introduction}

The problem of adapting topology based methods to
discrete data is an active field of research.
The set
of algorithmic tools for computing
topological properties in the setting of
digital imagery is relatively small. Mainly, Betti numbers, Euler
characteristic, skeletonization and  local
topological characterization. Recently, homology groups have also been computed in the
digital imagery  \cite{Lachaud}.

In order to enlarge this set, an algebraic-topological
representation for a given geometric object is developed in
\cite{GR05} in terms of a chain homotopy equivalence  (working with coefficients in a field).
These models allows to compute the (co)homology groups,
representative (co)cycles of the generators,
the cup product on
cohomology as well as the cohomological number $HB1$. Our
peculiar approach comes from the following main sources:
(a) Eilenberg-Mac Lane work on homology of simplicial sets
\cite{EM53}; (b) Effective Homology Theory  \cite{F87}.

In this paper, we extend the results given in \cite{GR05} to the
integer domain.  Algebraic-topological
representations of geometric objects
are expressed here only in terms of  particular
chain homotopies (called  AM-models).
Compared to previous works, these models also encode torsion groups.
Moreover, all the algorithms for
computing integer homology based on the matrix reduction method
to Smith normal form (for example
\cite{Mun84,Ago76,DSV01,Lachaud}) can be translated into our
setting without extra computational cost. Finally, we
successfully apply our computational algebraic-topological
approach to 3D  binary digital images. Moreover, AM-models for 3D binary digital images
can be   reused after voxel-set operations (union,
intersection, difference and inverse).


\section{Algebraic-topological Models for Integer Homology Computation}

In this section, we  deal with the concept of AM-model
(first given in \cite{GR01} and  \cite{GR06})  for the computation of all the integer homological
information of a
simplicial complex. In fact, we redefine this concept in simpler terms, what allows to
store the same information in less space.    

We first introduce some basic algebraic-topological
notions needed throughout the paper, which are extracted, mainly,
from \cite{Mun84}.
Let ${\bf Z}$ be the ground ring. Considering an ordering on a
vertex set $V$, a simplex with $q+1$ affinely independent
vertices $v_0<\cdots<v_q$ of $V$ is the convex hull of these
points, denoted by $\sigma^q=\langle v_0,\dots,v_q\rangle$.
The {\em dimension} of $\sigma^q$ is $q$.
 If $i<q$, an $i$--{\em face} of $\sigma^q$ is
 an $i$--simplex whose vertices belong to the set
 $\{v_0,\dots,v_q\}$.
  A {\em simplicial complex} $K$ is a collection of simplices such
 that every face of a simplex of $K$ is also a simplex of $K$ and the
 intersection of any two simplices of $K$ is either a face of both or empty. The {\em dimension of $K$} is the one of the
 highest dimensional simplex in $K$. $K^{(q)}$ denotes the set of all the $q$--simplices of $K$.

A {\em chain complex} ${\cal C}$ is a
sequence
$$\cdots\,\stackrel{d_{q+1}}{\rightarrow} {\cal C}_{q}\stackrel{d_q}{\rightarrow}
{\cal C}_{q-1}\stackrel{d_{q-1}}{\rightarrow}\,
\cdots \,\stackrel{d_1}{\rightarrow}
{\cal C}_0\stackrel{d_0}{\rightarrow}0\,.$$
of abelian groups ${\cal C}_q$ (called {\em $q$-chain groups}) and homomorphisms $d_q:{\cal C}_q\to {\cal C}_{q-1}$
(called the {\em differential}
of ${\cal C}$ in dimension $q$),
indexed with the integers,
such that $d_{q-1}d_q=0$ for all $q$.
In this paper, each group ${\cal C}_q$ is free of finite rank.
A chain complex ${\cal C}$
can be encoded as a couple $(C,d)$, where: (1) $C=\{C_q\}$
and, for each $q$, $C_q$ is a base of ${\cal C}_q$;
(2) $d=\{d_q\}$  and, for each $q$, $d_q$ is the differential of ${\cal C}$ in dimension $q$
with respect to the
bases $C_q$ and $C_{q-1}$.

Such an algebraic
structure can be associated to a given simplicial complex $K$, in
the following way: a $q$--{\em chain} $a^q$ is a finite sum $\sum
\lambda_i\sigma^q_i$ where $\lambda_i\in {\bf Z}$ and $\sigma^q_i\in
K^{(q)}$. The $q$--chains form  the
 {\em $q$th chain group} of $K$, denoted by ${\cal C}_q(K)$ (with $q\geq 0$).
  The {\em boundary} of a $q$--simplex $\sigma^q=\langle
v_0,\dots,v_q\rangle$ is the  $(q-1)$--chain $\partial_q
  (\sigma^q)=\sum_{i=0}^q(-1)^i\langle v_0,\dots,\hat{v}_i,\dots, v_q\rangle$,
   where
 $\hat{v}_i$ means that $v_i$ is omitted.
 By
 linearity, $\partial_q$ can be extended to
 $q$--chains. Then, the  chain complex  ${\cal C}(K)$ is the collection of chain groups
${\cal C}_q(K)$ connected by the boundary operators $\partial_q$.

Given a chain complex ${\cal C}=(C,d)$, a $q$--chain $a^q\in {\cal C}_q$ is called a
$q$--{\em cycle} if $d_q  (a^q)=0$.
If $a^q=  d_{q+1} (b^{q+1})$ for some $b^{q+1}\in {\cal C}_{q+1}$ then $a^q$
 is called a $q$--{\em boundary}.
 Denote the groups of $q$--cycles
 and $q$--boundaries by $Z_q$ and $B_q$ respectively.
  Define the {\em $q$th  homology group} to be the quotient group
 $Z_q/B_q$, denoted by ${\cal H}_q({\cal C})$.
We say that two $q$-cycles $a^q$ and $b^q$ are {\em homologous} if there exists
a $(q+1)$-chain $c^{q+1}$ such that $a^q=b^q+d_{q+1}(c^{q+1})$.
For each $q$, the integer $q$th homology group ${\cal H}_q({\cal C})$ is
a finitely generated abelian group
isomorphic to $F_q\oplus T_q$, where
$F_q={\bf Z}\oplus\cdots\oplus {\bf Z}$ and $T_q={\bf Z}/t_1^q\oplus \cdots\oplus {\bf Z}/t_{m_q}^q$
are the {\em free subgroup} and the {\em torsion
subgroup} of ${\cal H}_q({\cal C})$, respectively.
The numbers $t_1^q,\dots,t_{m_q}^q$ which satisfy that $t_1^q\geq 2$ and $t_1^q | t_2^q |\cdots | t_{m_q}^q$
are called the {\em torsion coefficients} of ${\cal H}_q({\cal C})$.
The rank of $F_q$,
denoted by $\beta_q$,  is called the $q$th {\em Betti number} of
${\cal C}$.
Intuitively,
 $\beta_0$ is the number of components of connected pieces, $\beta_1$ is the number of independent ``holes" and
 $\beta_2$ is the number of ``cavities".
 For all $q$, there exists a finite number of elements of ${\cal H}_q({\cal C})$
 from which we can deduce all ${\cal H}_q({\cal C})$ elements. These elements are called {\em homology generators} of
 dimension $q$.
 We say that $a^q$ is a {\em representative $q$--cycle} of a homology generator $\alpha^q$ of dimension $q$ if
 $\alpha^q=a^q+B_q$. We denote $\alpha^q=[a^q]$.
Finally, the homology of a simplicial complex $K$ is defined as the homology of  ${\cal C}(K)$.

 As far as the homology computation of a chain complex is concerned, the classical algorithm
for computing homology with coefficients in  ${\bf Z}$ is the
integer reduction algorithm \cite{Mun84}.
Given a chain complex ${\cal C}=(C,d)$, this algorithm consists in reducing the
matrix  of the boundary operator in each dimension  $q$, to its
Smith Normal Form (SNF), relative to some bases $\{a^q_1,\dots
a^q_{m_q}\}$ of ${\cal C}_q$ and $\{e^{q-1}_1,\dots, e^{q-1}_{m_{q-1}}\}$ of  ${\cal C}_{q-1}$ such
that
 for some $t_q,\ell_q,s_{q-1}$ where $1\leq s_{q-1}\leq m_{q-1}$ and
 $1\leq t_q\leq\ell_q\leq$ min$(m_q,s_{q-1})$,
\begin{itemize}
\item $\partial_{q-1}(e^{q-1}_i)=0$ for $1\leq i\leq s_{q-1}$ and   $\partial_{q-1}(e^{q-1}_i)\neq 0$
for  $s_{q-1}< i\leq m_{q-1}$.
\item $\partial_q(a^q_i)=e^{q-1}_i$, for $1\leq i\leq t_q$;
\item $\partial_q(a^q_i)=\lambda^q_i e^{q-1}_i$, where $\lambda^q_i\in{\bf Z}$ and  $\lambda^q_i\geq 2$
for $t_q< i\leq \ell_q$;
\item and
$\partial_q(a^q_i)=0$ for $\ell_q< i\leq m_q$.
\end{itemize}
In this case,
$$F_{q-1}={\bf Z}\oplus \stackrel{\scst s_{q-1}-\ell_q}\cdots\oplus{\bf Z}\qquad \mbox{ and }\qquad
T_{q-1}={\bf Z}/\lambda^q_{t_q+1}\oplus\cdots\oplus{\bf Z}/\lambda^q_{\ell_q}\,.$$
 Moreover,
$\{e^{q-1}_{\ell_q+1},\dots,e^{q-1}_{s_{q-1}}\}$
and $\{e^{q-1}_{t_q+1},\dots,e^{q-1}_{\ell_q}\}$
are sets of representative cycles of the generators of $F_{q-1}$ and
 $T_{q-1}$, respectively.

In \cite{Lachaud}, an
algorithm improving the efficiency of this classical integer
reduction algorithm  is described. Their technique is
mainly based on the results of \cite{DSV01}, in which a matrix
reduction to integer SNF is determined in an efficient way. We
can take advantage of these improvements in
the algorithms described here without additional computational cost.


 A {\em chain contraction}  \cite{McL95} $(f,g,\phi)$ of a chain complex ${\cal C}=(C,d)$ to
a chain complex ${\cal C'}=(C',d')$ is a set of three
 homomorphisms $f=\{f_q: {\cal C}_q\to {\cal C}'_q\}$, $g=\{g_q: {\cal C}'_q\to {\cal C}_q\}$ and
 $\phi=\{\phi_q: {\cal C}_q\to {\cal C}_{q+1}\}$ such that for each $q$:
 \begin{itemize}
 \item  $f_{q-1}d_q=d'_qf_q$ and
$d_qg_q=g_{q-1}d'_q$.
\item  $f_q g_q$ is the identity map of ${\cal C'}_q$;
\item $\phi_{q-1}  d_q+d_{q+1} \phi_q=id_q-g_q f_q$, that is,
$\phi$ is
a chain homotopy of the identity map $id=\{id_q\}$ of ${\cal C}$ to $gf$.
\end{itemize}
 In this case,
${\cal C'}$ has fewer or the same number of generators than
${\cal C}$ while ${\cal C}$ and ${\cal C'}$ have isomorphic
homology groups \cite[p. 73]{Mun84}.

A translation  of the
 integer reduction algorithm in terms of  chain contractions  has been made in \cite{GR01,GR06}.
 In those papers an AM-model for a given simplicial complex $K$ was defined as a chain contraction of ${\cal C}(K)$
to a chain complex ${\cal M}$ such that the SNF, $A_q$, of the matrix of the differential of ${\cal M}$
in each dimension $q$ satisfies that any non-null entry of $A_q$ is greater than $1$.
Here we go further and define an AM-model for a given simplicial complex only in terms of the chain homotopy $\phi$.
We prove that it is possible to recover the other homomorphisms $f$ and $g$ and the chain complex ${\cal M}$.
Moreover, our  strategy
outperforms the previous algorithms for computing integer homology
  in several points: (1) cohomological features can be computed;
 (2) we can control the topological changes after addition or
deletion of simplices.

\begin{definition}
An {\em AM-model} for a simplicial complex $K$ is a couple
$(C, \phi)$ such that:
\begin{itemize}
\item  $C=\{C_q\}$ and, for each $q$, $C_q$ is a base of ${\cal C}_q(K)$;
\item $\phi=\{\phi_q\}$ and, for each $q$, $\phi_q:{\cal C}_q(K)\to {\cal C}_{q+1}(K)$
is a homomorphism satisfying that
$\phi_{q+1}\phi_{q}=0$ and $\phi_q \partial_{q+1}\phi_{q}=\phi_{q}$;
\item the chain complex
${\cal M}=(M,d)$
(such that $M_q=$ Im $\pi_q$,
 $\pi_q=id_q-\phi_{q-1} \partial_q
-\partial_{q+1} \phi_q$ and  $d_q=\partial_q|_{\scst {\cal M}_q}$), satisfies,  for each $q$, that
 any non-null entry of the
SNF of $d_q$ is greater than $1$.
\end{itemize}
\end{definition}

\begin{thm} Given an AM-model $(C,\phi)$ for a simplicial complex $K$, we can define a chain contraction
$(f,g,\phi)$ of ${\cal C}(K)$ to a chain complex
${\cal M}$ such that any non-null entry of the SNF of the matrix of the differential of ${\cal M}$, for each $q$,
is greater than $1$.
In particular, if the homology of $K$ is free
then ${\cal M}$  is isomorphic  to
${\cal H}(K)$.
\end{thm}

\noindent {\bf Proof.}
We only have to define $f$ as $\{\pi_q\}$ where, for each $q$, $\pi_q=id_q-\phi_{q-1} \partial_q
-\partial_{q+1} \phi_q$, and $g$ as the inclusion. Let us see that
$(f,g,\phi)$ is a chain contraction. It is clear that, for each $q$,
$g_qf_q=id_q-\phi_{q-1}\partial_q-\partial_{q+1}\phi_q$.
On the other hand, let $a^q\in {\cal M}_q$. There exists  $b^q\in {\cal C}_q(K)$ such that
$a^q=\pi_q(b^q)=b^q-\phi_{q-1}\partial_q(b^q)-\partial_{q+1}\phi_q(b^q)$.
Then, $f_qg_q(a^q)=\pi_q(a^q)=b^q-\phi_{q-1}\partial_q(b^q)-\partial_{q+1}\phi_q(b^q)
-\phi_{q-1}\partial_q(b^q)+\phi_{q-1}\partial_q\phi_{q-1}\partial_q(b^q)
+\phi_{q-1}\partial_q\partial_{q+1}\phi_q(b^q)-\partial_{q+1}\phi_q(b^q)+\partial_{q+1}\phi_q\phi_{q-1}\partial_q(b^q)
+\partial_{q+1}\phi_q\partial_{q+1}\phi_q(b^q)=a^q$.
Second, $f_{q-1}\partial_q=\partial_q-\partial_q\phi_{q-1}\partial_q=\partial_q f_q$. Finally,
if $a^q\in {\cal M}_q$ then $\partial_q(a^q)\in{\cal M}_q$. Therefore
$g_{q-1}\partial_q=\partial_q g_q$.
\qed

Given a simplicial complex $K$, it is possible to define different
AM-models for $K$ since  the chain homotopy $\phi$ and the chain complex ${\cal M}$
are not unique. In the following lemma, we show how to compute a new AM-model from a previous one.

\begin{lem}\label{lema1} Let $(C,\phi)$ be an AM-model for a simplicial complex $K$ and let $h^q\in {\cal C}_q(K)$.
\begin{itemize}
\item If $h^q\in {\cal M}_q$ then $\phi_q(h^q)=0$; and if  $\partial_q(h^q)=0$ then $\pi_q(h^q)=h^q$.
\item If $x^q\in {\cal C}_q(K)$ and $h^q\in{\cal M}_q$ such that  $\partial_q(x^q)=0$
and $\pi_q(x^q)=h^q$ then $[x^q]=[h^q]$ and we can define a new AM-model
$(C,\phi')$ as $\phi'_q(x^q):=\phi_q(h^q)$, $\phi'_q(h^q):=\phi_q(x^q)$ and $\phi':=\phi$ for the rest. In this case,
$\pi'_q(x^q)=(id_q-\phi'_{q-1}\partial_q-\partial_{q+1}\phi'_q)(x^q)=x^q$ and $\pi'_q(h^q)=x^q$.
\end{itemize}
\end{lem}

\begin{alg}\label{ammodel} Computing an AM-model for a Finite Simplicial Complex.
\begin{tabbing}
{\sc Input: }\={\tt A simplicial complex $K$ of dimension $n$.}\\
{\tt For } \= {\tt $q=1$ to $q=n$ do}\\
\> {\tt  reduce the matrix  of $\partial_q$ to its SNF relative to some bases}\\
\>{\tt  $\{a^q_1,\dots,a^q_{m_q}\}$ of ${\cal C}_{q}(K)$
and $\{e^{q-1}_1,\dots,e^{q-1}_{m_{q-1}}\}$ of ${\cal C}_{q-1}(K)$ such that}\\
\> {\tt for some }\= {\tt $s_{q-1}$ where $1\leq s_{q-1}\leq m_{q-1})$:}\\
\> \>{\tt $\partial_{q-1}(e^{q-1}_i)=0$, for $1\leq i\leq s_{q-1}$}\\
\>\> {\tt  and $\partial_{q-1}(e^{q-1}_i)\neq 0$,
for $s_{q-1}<i\leq m_{q-1}$;}\\
\> {\tt for some }\= {\tt $t_q,\ell_q$ where $1\leq t_q\leq\ell_q\leq$ min$(m_q,s_{q-1})$:}\\
\> \>{\tt $\partial_q(a^q_i)=e^{q-1}_i$, for $1\leq i\leq t_q$;}\\
\>\>{\tt $\partial_q(a^q_i)=\lambda^q_i e^{q-1}_i$, $\lambda^q_i\in{\bf Z}$, $\lambda^q_i\geq 2$
for $t_q< i\leq \ell_q$;}\\
\>\>{\tt   and
$\partial_q(a^q_i)=0$ for $\ell_q< i\leq m_q$. }\\
 \>{\tt Define }\= {\tt   $C_{q-1}:= \{e^{q-1}_1, \ldots, e^{q-1}_{m_{q-1}}\}$,
 $C_{q} := \{a^q_1, \ldots, a^q_{m_q}\}$,}\\
 \>\> {\tt  $\phi_{q-1}(e^{q-1}_i):= a^q_i$ {\tt for} $1 \leq i \leq t_q$,}\\
 \>\> {\tt
 $\phi_{q-1}(e^{q-1}_i):= 0$ {\tt for} $t_q< i \leq s_{q-1}$}\\
 \>\> {\tt  and
$\phi_q(a^q_i):= 0$ {\tt for} $1\leq  i \leq t_q$.}\\
{\sc Output:} \= {\tt The couple $(C,\phi)$.}
\end{tabbing}
\end{alg}

Notice that, for $q=k$, $\phi_k(a^k_i)$ is defined for $1\leq i \leq t_k$.
After, $\phi_k$ is again defined when $q=k+1$, but not for the previous elements $a^k_i$,
 since $a^k_i\not\in$ Im $\partial_{k+1}$, for $1\leq i \leq t_k$.

\begin{thm}
Let $K$ be a finite simplicial complex of dimension $n$.
The output of Algorithm \ref{ammodel},  $(C,\phi)$, defines an AM-model for $K$.
Moreover, integer homology generators and representative cycles of these generators can be directly obtained
from ${\cal M}$.
\end{thm}

\noindent{\bf Proof.}
First of all, let us see that for each $q$, if  $x^q\in C_q$, then $\pi_q(x^q)=0$ or $\pi_q(x^q)=x^q$,
where $\pi_q=id_q-\phi_{q-1} \partial_q
-\partial_{q+1} \phi_q$.
Since the matrix of the differential in dimension $q$ with respect to the base $C_q$ coincides with its SNF, then for each
$x^q\in C_q$:
\begin{itemize}
\item If $\partial_q(x^q)=y^{q-1}$ for some $y^{q-1}\in C_{q-1}$,
 then $\phi_{q-1}(y^{q-1})=x^q$ and $\phi_{q}(x^q)=0$. Therefore $\pi_q(x^q)=0$.
\item If $\partial_q(x^q)=\lambda y^{q-1}$, for some $y^{q-1}\in C_{q-1}$ and $\lambda\in {\bf Z}$,  $\lambda\geq 2$,
then $\phi_{q-1}(y^{q-1})=0$
and $\phi_q(x^{q})=0$. Therefore, $\pi_q(x^q)=x^q$.
\item If $\partial_q(x^q)=0$ and there exist $z^{q+1}\in C_{q+1}$ such that $\partial_{q+1}(z^{q+1})=x^q$
then $\phi_q(x^q)=z^{q+1}$ and $\pi_q(x^{q})=0$.
\item If $\partial_q(x^q)=0$ and there exist $z^{q+1}\in C_{q+1}$ and $\lambda\in {\bf Z}$,  $\lambda\geq 2$, such that
$\partial_{q+1}(z^{q+1})=\lambda x^{q}$, then
$\phi_q(x^q)=0$ and $\pi_q(x^q)=x^q$.
\item If $\partial_q(x^q)=0$ and there is neither $\lambda\in {\bf Z}$, $\lambda\neq 0$, nor $z^{q+1}\in C_{q+1}$,
such that $\partial_{q+1}(z^{q+1})=\lambda x^q$, then
 $\phi_q(x^q)=0$ and $\pi_q(x^q)=x^q$.
\end{itemize}
Therefore, a base of ${\cal M}$ in each dimension $q$ is the set $M_q=\{x^q:\; x^q\in C_q$ and $\pi_q(x^q)=x^q\}$.
Now, for each $q$, let $\{x^q_1,\dots,x^q_{m_q}\}$ be the elements of $M_q$
and $\{y^{q-1}_1,\dots,y^{q-1}_{m_{q-1}}\}$  the elements of $M_{q-1}$.
For some $s_{q-1}$, $1\leq s_{q-1}\leq m_{q-1}$,
 $\partial_{q-1}(y^{q-1}_i)=0$ for $1\leq i\leq s_{q-1}$
and  $\partial_{q-1}(y^{q-1}_i)\neq 0$ for $s_{q-1}< i\leq m_{q-1}$.
For some $\ell_q$ where  $1\leq\ell_q\leq$ min$(m_q,s_{q-1})$,
\begin{itemize}
\item $\partial_q(x^q_i)=\lambda^q_i y^{q-1}_i$, where $\lambda^q_i\in{\bf Z}$ and  $\lambda^q_i\geq 2$
for $1\leq i\leq \ell_q$;
\item $\partial_q(a^q_i)=0$ for $\ell_q< i\leq m_q$.
\end{itemize}
In this case,
$$F_{q-1}={\bf Z}\oplus \stackrel{\scst s_{q-1}-\ell_q}\cdots\oplus{\bf Z}\qquad \mbox{ and }\qquad
T_{q-1}={\bf Z}/\lambda^q_{1}\oplus\cdots\oplus{\bf Z}/\lambda^q_{\ell_q}\,.$$
 Moreover,
$\{y^{q-1}_{\ell_q+1},\dots,y^{q-1}_{s_{q-1}}\}$
and $\{y^{q-1}_{1},\dots,y^{q-1}_{\ell_q}\}$
are sets of representative cycles of the generators of $F_{q-1}$ and
 $T_{q-1}$, respectively.
\qed

\begin{figure}[t!]
\centerline{\includegraphics[width=7cm]{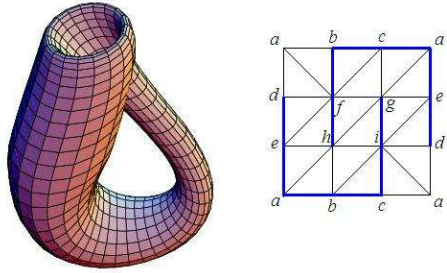}}
 \caption{The Klein bottle and a triangulation of it.}
\end{figure}

\begin{exmp}\label{ex1} Consider the  simplicial complex $K$ in Figure 1 whose underlying
space is the Klein bottle \cite[p. 283]{Mun84}. Applying Algorithm \ref{ammodel},
we obtain that the vertex $\langle a \rangle$
is an element of $C_0$. The rest of the elements of $C_0$ are
the boundaries of the $1$-simplices marked in blue in Figure 1.
These $1$-simplices  are also elements of $C_1$. Denote by $x$
one of these $1$-simplices. The cycles
$\alpha_1:=\langle a,d\rangle+\langle d,e\rangle-\langle
a,e\rangle$  and
$\alpha_2:=\langle
a,b\rangle+\langle b,c\rangle-\langle a,c\rangle$  are also elements of $C_1$. The rest of the elements of $C_1$ are
the boundaries of all the $2$-simplices except for $\langle
f,g,h\rangle$. These $2$-simplices belong to  $C_2$. Denote by
$y$ one of these $2$-simplices. The $2$-chain
consisting in the sum of all the triangles in $K$,
$\gamma:=-\langle a,b,f\rangle-\langle b,c,f\rangle+\langle
a,c,g\rangle-\langle a,e,g\rangle+\langle e,g,i\rangle- \langle
e,d,i\rangle+\langle c,d,i\rangle-\langle a,c,d\rangle+\langle
b,c,i\rangle+\langle a,b,h\rangle- \langle a,e,h\rangle-\langle
e,d,f\rangle+\langle a,d,f\rangle+\langle c,f,g\rangle-\langle
f,g,h\rangle- \langle h,g,i\rangle-\langle b,h,i\rangle$, is also an
element of $C_2$. The image of $\partial$, $\phi$ and $\pi$ on
$C$  and the SNF of the matrix of the differential in dimension $1$ and $2$ are given
below:
$$\begin{array}{c|cccccccc}
C&\;\langle a\rangle\;&\;\partial_1(x)\;&\;x\;&\;\alpha_1\;&\;\alpha_2\;&\;\partial_2(y)\;&\;y\;&\;\gamma\;\\
\hline
\partial &0&0&\;\partial_1(x)\;&0&0&0&\;\partial_2(y)\;&\;2\alpha_1\;\\
\phi& 0&x&0&0&0&y&0&0\\
\pi& \langle a\rangle\;&0&0&\;\alpha_1\;&\;\alpha_2\;&0&0&\;\gamma\;
\end{array}$$
$$\begin{array}{c|cccc}
\partial_1& x&\partial_2(y)&\alpha_1&\alpha_2\\
\hline
\partial_1(x)& 1&0&0&0\\
\langle a\rangle&0&0&0&0
\end{array}
\qquad\qquad
\begin{array}{c|cc}
\partial_2& y&\gamma\\
\hline
\partial_2(y)& 1&0\\
\alpha_1& 0&2\\
\alpha_2&0&0
\end{array}
$$
Therefore, $M_0=\{\langle a\rangle\}$,
$M_1=\{\alpha_1,\alpha_2\}$ and $M_2=\{\gamma\}$, $\partial|_{{\cal M}_0}(\langle
a\rangle)=0$, $\partial|_{{\cal M}_1}(\alpha_1)=0$, $\partial|_{{\cal M}_1}(\alpha_2)=0$ and
$\partial|_{{\cal M}_2}(\gamma)=2\alpha_1$. We obtain  that
${\cal H}_0(K)\simeq {\bf Z}$, ${\cal H}_1(K)\simeq {\bf Z}\oplus {\bf Z}/2$ and representative cycles of the homology generators are
$\langle a\rangle$ for ${\cal H}_0(K)$, $\alpha_2$ for the free part of
${\cal H}_1(K)$ and $\alpha_1$ for the torsion part.
\end{exmp}

\section{Extracting Integer Cohomology Information from AM-models}

In this section, we extend the work done in  \cite{GR05} (with coefficients in a field)
for computing   cohomology (the dual notion of homology) over the integer domain.
Working with coefficients in a field, homology groups are free and
 isomorphic to cohomology groups.
Nevertheless, integer homology and cohomology
can have a torsion part and they are not isomorphic, in general.
The cohomology groups
  have an additional multiplicative structure, the cup product,
 from which we can derive
finer invariants than homology.

Let  ${\cal C}=(C,d)$ be a chain complex. The
{\em cochain complex } ${\cal C}^*=(C^*,\delta)$ in each dimension $q$ is the
group of $q$--{\em cochains}
 with coefficients in ${\bf Z}$,
${\cal C}^q=\{c:{\cal C}_q\ra {\bf Z}$  such that
$c$ is a homomorphism$\}$.
If $C_q=\{a_1,\dots,a_{m_q}\}$  is  a base of
 ${\cal C}_q$ then a base of ${\cal C}^q$ is $C^q=\{a_1^*,\dots,a_{m_q}^*\}$,
 where $a^*_i: {\cal C}_q\to {\bf Z}$ is given by $a^*_i(a_i)=1$ and $a^*_i(a_j)=0$ for
 $1\leq i,j\leq m_q$ and $j\neq i$.
For all $q$, the differential  $d_{q+1}$ on ${\cal C}_{q+1}$ induces the {\em codifferential}
$\delta^{q}: {\cal C}^q\ra {\cal C}^{q+1}$ via $\delta^{q} (c)=c d_{q+1}$,
so that $\delta^q$ increases the
dimension by one.
Define $Z^q$ to be the kernel of $\delta^q$ and
$B^{q+1}$ to be its image.
These groups are called the group of $q$--{\em cocycles} and $q+1$--{\em coboundaries}, respectively.
Define
the $q$th {\em cohomology group},
${\cal H}^q({\cal C})=Z^q/B^q$  for $q\geq 0$.
For each $q$, the integer $q$th cohomology group ${\cal H}^q({\cal C})$ is
a finitely generated abelian group
isomorphic to $F^q\oplus T^q$, where
$F^q$ and $T^q$ are the {\em free subgroup} and the {\em torsion
subgroup} of ${\cal H}^q({\cal C})$, respectively. The rank of $F^q$ coincides with the rank of $F_q$ for each $q$.

\begin{thm}
Let $(C,\phi)$ be the output of Algorithm \ref{ammodel} for a given finite simplicial complex $K$.
The integer cohomology of $K$ and representative cocycles of integer cohomology generators  can be
directly obtained from ${\cal M}=$ im $\pi$, where $\pi=id-\phi\partial-\partial\phi$.
\end{thm}

\noindent{\bf Proof.}
A base of ${\cal M}$ is the set $M=\{x:\; x\in C$ and $\pi(x)=x\}$.
Now, for each $q$, let $\{x_1,\dots,x_{m_q}\}$ be the elements of $M_q$
and $\{y_1,\dots,y_{m_{q-1}}\}$  the elements of $M_{q-1}$.
For some $s_{q-1}$, $1\leq s_{q-1}\leq m_{q-1}$,
 $\partial_{q-1}(y_i)=0$ for $1\leq i\leq s_{q-1}$
and  $\partial_{q-1}(y_i)\neq 0$ for $s_{q-1}< i\leq m_{q-1}$.
For some $\ell_q$ where  $1\leq\ell_q\leq$ min$(m_q,s_{q-1})$,
$\partial_q(x_i)=\lambda_i y_i$, where $\lambda_i\in{\bf Z}$ and  $\lambda_i\geq 2$
for $1\leq i\leq \ell_q$;
and
$\partial_q(x_i)=0$ for $\ell_q< i\leq m_q$.
In this case,
$$F^{q-1}={\bf Z}\oplus \stackrel{\scst s_{q-1}-\ell_q}\cdots\oplus{\bf Z}\qquad \mbox{ and }\qquad
T^{q}={\bf Z}/\lambda_{1}\oplus\cdots\oplus{\bf Z}/\lambda_{\ell_q}\,.$$
 Moreover,
$\{y^*_{\ell_q+1},\dots,y^*_{s_{q-1}}\}$
and $\{x^*_{1},\dots,x^*_{\ell_q}\}$
are sets of representative cocycles of the generators of $F^{q-1}$ and
 $T^{q}$, respectively.
\qed

\begin{exmp} Consider the AM-model $(C,\phi)$ obtained in
Example \ref{ex1} for the simplicial complex $K$ whose underlying space is the Klein bottle.
Starting from the chain complex ${\cal M}$ whose base is
$\{\langle a\rangle, \alpha_1,\alpha_2,\gamma\}$ and differential
$\partial|_{\scst {\cal M}}$, we construct in a straightforward way the cochain complex ${\cal M}^*$ whose base is
$\{\langle a\rangle^*, \alpha_1^*,\alpha_2^*,\gamma^*\}$ and codifferential $\delta$ given by:
$\delta_0(\langle a\rangle^*)=\langle a\rangle^*\partial|_{\scst {\cal M}_1}=0$,
$\delta_1(\alpha_1^*)=\alpha_1^*\partial|_{\scst {\cal M}_2}=0$,
$\delta_1(\alpha_2^*)=\alpha_2^*\partial|_{\scst {\cal M}_2}=2\gamma^*$,
$\delta_2(\gamma^*)=\gamma^*\partial|_{\scst {\cal M}_3}=0$.

Therefore we obtain  that ${\cal H}^0(K)\simeq {\bf Z}$, ${\cal H}^1(K)\simeq {\bf Z}$ and ${\cal H}^2(K)\simeq {\bf Z}/2$;
 and the representative cocycles  are:
$\langle a\rangle^*$ in dimension $0$, $\alpha^*_1$ in dimension $1$
and $\gamma^*$ in dimension $2$.
\end{exmp}

The cochain complex ${\cal C}^*(K)$ is a ring with
the {\em cup product} $\smile: {\cal C}^p(K)\ti {\cal C}^q(K)\ra
{\cal C}^{p+q}(K)$ given by:
$$\;(c\smile c')(\langle v_0,\dots,v_{p+q}\rangle)=
c (\langle
v_{0},\dots,v_{p}\rangle)\cdot c'(\langle
v_p,\dots,v_{p+q}\rangle)\,.$$
This product induces an operation
$\smile: {\cal H}^p(K)\ti {\cal H}^q(K)\ra {\cal H}^{p+q}(K)$, via $[c]\smile [c']=[c\smile c']$,
that is bilinear,
associative, commutative up to a sign,
independent of the ordering of the vertices of $K$
and homotopy-type invariant \cite[p. 289]{Mun84}.

Working with coefficients in ${\bf Z}/2$, a new cohomology invariant called $HB1$ is obtained in \cite{GR05}.
The idea is to put into
a matrix form the multiplication table of the cup product of cohomology generators of dimension $1$.
The following algorithm
compute $HB1$ working with integer coefficients.
Assume that
an AM-model
for $K$, $(C,\phi)$, is computed using Algorithm \ref{ammodel}. Then, for each $q$,
 $M_q=\{x:\;x\in C_q$ and $\pi_q(x)=x\}$ is a base of
${\cal M}_q=$ im $\pi_q$, where $\pi_q=id_q-\phi_q\partial_q-\partial_q\phi_q$. Let $M^q=\{x^*:\;x\in M_q\}$ where
$x^*:{\cal M}_q\to {\bf Z}$ is such that $x^*(x)=1$ and $x^*(z)=0$ for $z\in M_q$ and $z\neq x$.
Suppose that there are $r_i$ elements in $C_i$
and $s_i$ cocycles  in $M^i$, $i=1,2$. Then
the following algorithm computes the cohomological invariant $HB1$, working with integer coefficients,
in ${\cal O}(r_1^2s_1^2r_2s_2)$.

\begin{alg} Algorithm for computing the cohomological invariant $HB1$.
\begin{tabbing}
{\sc Input: }\= {\tt An AM-model $(C,\phi)$ for a simplicial complex $K$ computed}\\
\> {\tt  using Algorithm \ref{ammodel}.}\\
{\tt Let $M^*=\{x^*:\; x\in C$ and $\pi(x)=x\}$.}\\
{\tt Let $\{\alpha^*_1,\dots,\alpha^*_p\}$ and  $\{\gamma^*_1,\dots,\gamma^*_m\}$ be the sets of $1$ and $2$-cocycles}\\
{\tt  in $M^*$, respectively.}\\
 {\tt For }\= {\tt  $i=1$ to $p$ do}\\
\> {\tt For }\= {\tt  $j=i$ to $p$ do}\\
\>\> {\tt For }\= {\tt  $k=1$ to $m$ do}\\
\>\>\>{\tt   $b_{((i,j),k)}:=
(\alpha^*_i \smile \alpha^*_j )(\gamma_{k})$.}\\
{\tt $HB1:=$  the rank of the 2D matrix of integers $(b_{((i,j),k)})$.}\\
{\sc Output: }\={\tt The integer $HB1$.}
\end{tabbing}
\end{alg}

The implementation of the algorithm described above
 has already been made  \cite{javier}. We have tested it on several 3D objects.
We give here an example of the computation of the cohomology, representative cocycles of cohomology generators
 and the invariant $HB1$.

 \begin{exmp}
Consider the simplicial complex  $T$ whose underlying space is showed in Figure 2 (on the left).
It consists in
 $11847$ simplices.
The running time for computing an AM-model  for $T$ and the homology of $T$ using a
Pentium 4, 3.2 GHz, 1Gb RAM is 2 seconds.
We obtain that $\beta_0=1$, $\beta_1=4$ and $\beta_2=3$.
 The running time for computing  the cup product is $1.5$ seconds. In Figure 2 (on the center),
 the
$1$ and $2$-simplices, on which the representative cocycles are non-null, are drawn. The table on the
right  of Figure 2 shows the results of the cup product of any two cohomology generators of dimension $1$.
Finally,
$HB1=2$.
\end{exmp}

\begin{figure}[t!]
\centerline{\includegraphics[width=14cm]{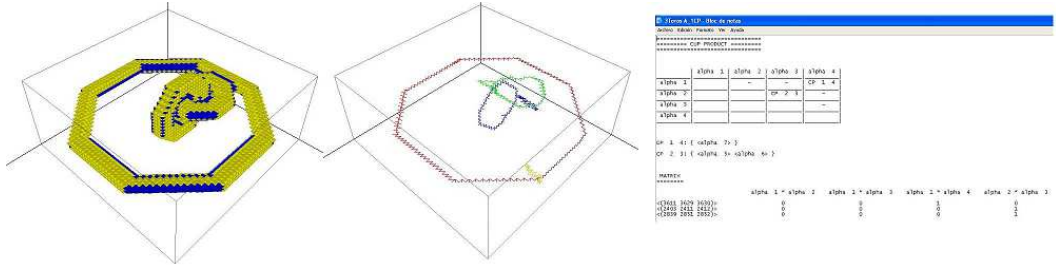}}
\caption{The simplicial complex $T$, representative cocycles of
the generators of ${\cal H}^1(T)$ and the multiplication table of the cup product.}
\end{figure}

\section{AM-models for 3D Digital Images}

An important issue in Digital Volume Processing is to design
efficient algorithms for analysis and processing in grids such as
the face-centered cubic (fcc) and the body-centered cubic (bcc)
grids \cite{Herman}, since it is very easy to obtain data
structures for them.
 The bcc and the fcc grid are the generalizations to 3D of the two-dimensional hexagonal
grid.
In the bcc grid, the voxels consist of truncated octahedra, and in the fcc grid, the voxels consist of
rhombic dodecahedra. They are better approximations of Euclidean balls than the cube.

A 3D digital image ${\cal I}$ is encoded as a tuple
$(V,I,b,w)$, where $V$ is the set of grid points in a 3D
grid, $I$ is the set of black points and $b$ (resp. $w$)
determines the neighborhood relations between black points (resp.
white points) in the grid. A bcc grid is equivalent to a grid
${\mathcal V}$ in which the grid points are those
$(x_1,x_2,x_3)\in {\bf Z}^3$ such that $x_1\equiv x_2\equiv x_3$
(mod $2$) (see \cite{Kov84}). The only Voronoi adjacency relation
on ${\mathcal V}$ is the $14$--adjacency. Using this
adjacency, it is straightforward to associate to a digital image
${\cal I}=({\mathcal V},I,14,14)$  a unique simplicial complex
$K(I)$ (up to isomorphism) with the same topological information
as $I$. It is called the {\em simplicial representation of $I$}.
The $i$--simplices of $K(I)$ ($i\in \{0,1,2,3\}$) are constituted
by the different sets of $i$ mutually $14$--neighbor points in
$I$. Since an isomorphism of digital images is equivalent to a
simplicial homeomorphism of the corresponding simplicial
representations, we define the (co)homology of  $I$ as the (co)homology
of $K(I)$. Moreover, we define an {\em AM-model} for ${\cal I}$
as an AM-model for its simplicial representation $K(I)$,
$(C_{\scst I},\phi_{\scst I})$, where $C_{\scst I}$ is a base
for ${\cal C}(K(I))$. Therefore, the simplicial complexes considered in
this section are embedded in ${\bf R}^3$, then their homology
groups  vanish for dimensions greater than $3$ and they are
torsion--free for dimensions $0$, $1$ and $2$ (see
\cite[ch.10]{AH35}). Moreover, the value of all the possible
non-null entries of the SNF of the matrix of the differential of
${\cal C}(K)$ (where $K$ is a simplicial complex embedded in ${\bf
R}^3$) in each dimension must be $1$; and an AM-model $(C,\phi)$
for $K$ satisfies that the chain complex ${\cal M}=$ im
$\pi=id-\phi\partial-\partial\phi$ is isomorphic to the homology
of $K$.

All the algorithms explained in this paper have been implemented
 \cite{javier} using as a grid the set of points with integer coordinates
 in the Euclidean $3$--space ${\bf Z}^3$
and the $14$--adjacency by which the neighbors of a grid point
(black or white) with integer coordinates $(x_1,x_2,x_3)$ are:
$(x_1\pm 1,x_2,x_3)$, $(x_1,x_2\pm 1,x_3)$, $(x_1,x_2,x_3\pm 1)$,
$(x_1+1,x_2-1,x_3)$, $(x_1-1,x_2+1,x_3)$, $ (x_1+1,x_2,x_3-1)$,
$(x_1-1,x_2,x_3+1)$, $(x_1,x_2+1,x_3-1)$, $(x_1,x_2-1,x_3+1)$,
$(x_1+1,x_2+1,x_3-1)$, $ (x_1-1,x_2-1,x_3+1)$. The digital spaces $({\bf Z}^3, 14,14)$  and $({\cal
V},14,14)$ are isomorphic: a grid point $(x_1,x_2,x_3)$ of $({\bf
Z}^3, 14,14)$ can be associated to
 the point $(x_1+x_2+2x_3,-x_1+x_2,-x_1-x_2)$ of $({\cal V},14,14)$.

\begin{figure}[t!]
\centerline{\includegraphics[width=8.5cm]{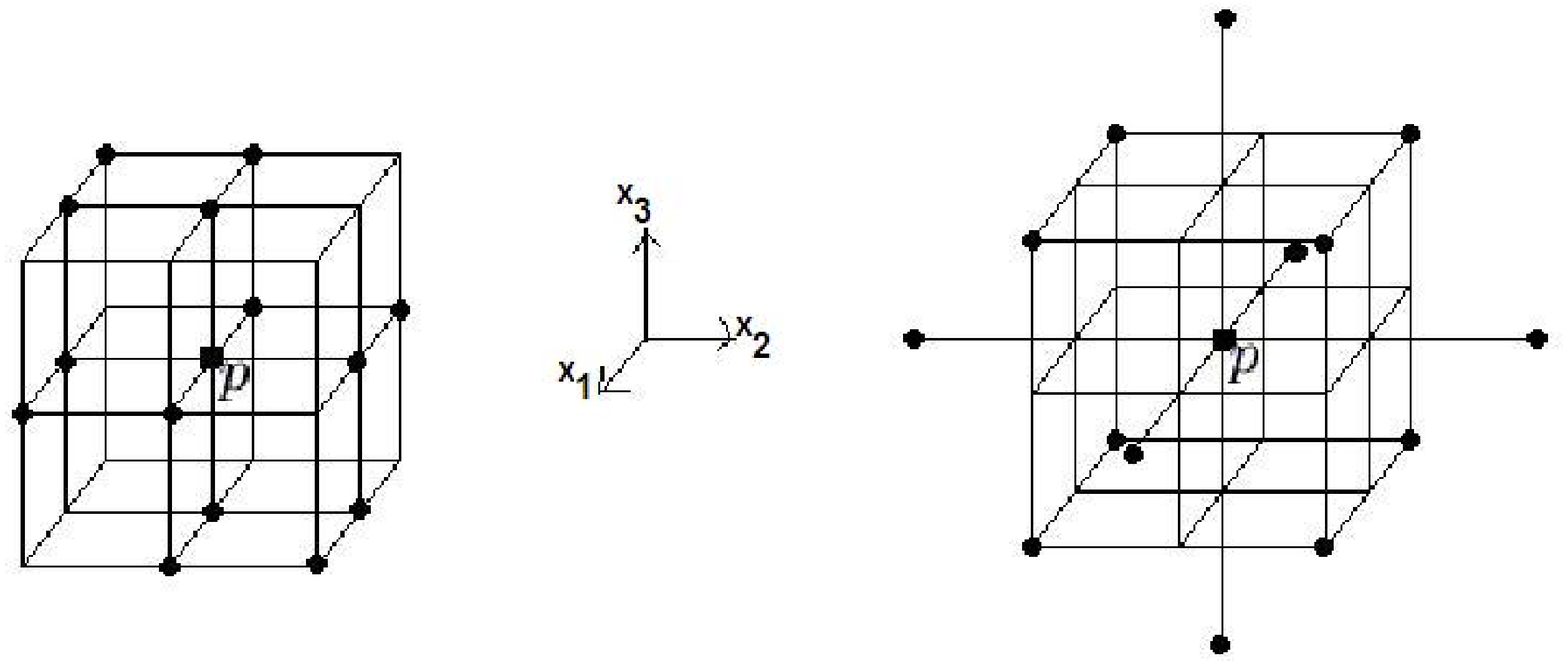}}
 \caption{The $14$--neighbors of a  point $p\in {\bf Z}^3$  (on the left) and $p\in {\cal V}$ (on the right).}
\end{figure}

In order to compute an AM-model for a digital image ${\cal I}=({\bf Z}^3,I,14,14)$,
we  take advantage of the particular structure of
$K(I)$ in the way that we  consider as the input of Algorithm \ref{ammodel} the following special initial base:
\begin{itemize}
\item In dimension $0$, it consists in the set of
the vertices of $K(I)$, except for the vertices $\langle(x_1,x_2,x_3)\rangle$ such that $x_3$ is odd,
 which are replaced by
the boundary of the $1$-chains $a=\langle (x_1,x_2,x_3-1),(x_1,x_2,x_3)\rangle$, if $a\in  K(I)$.
\item
In dimension $1$, it consists in the set of
all the edges in $K(I)$, except for the edges of the form
$\langle (x_1,x_2,x_3),(x_1,x_2+1,x_3)\rangle$,
which are replaced by
the boundary of the $2$-chains
 $b=\langle (x_1,x_2,x_3),(x_1,x_2,x_3+1),(x_1,x_2+1,x_3)\rangle$ if $b\in  K(I)$.
\item
In dimension $2$, it consists in the set of
all the triangles in $K(I)$, except for the triangles
of the form $\langle (x_1-1,x_2+1,x_3),(x_1,x_2,x_3),(x_1,x_2+1,x_3) \rangle$,
which are replaced by the boundary of the $3$-chains
$c=\langle (x_1-1,x_2,x_3+1),(x_1-1,x_2+1,x_3),(x_1,x_2,x_3),(x_1,x_2+1,x_3) \rangle$, if $c\in K(I)$.
\item
In dimension $3$, it consists in the set of all the tetrahedra in $K(I)$.
\end{itemize}
We then reduce the matrix of $\partial_q$ relative to this base to its SNF and it holds that we do not have to modify the
rows and columns
corresponding to the
chains $a$, $b$, $c$, $\partial(a)$, $\partial(b)$ and $\partial(c)$.

In the following table we present the running time for computing
AM-models for the 3D digital images showed in
Figure 4 using the program developed in \cite{javier}.

\begin{center}
\begin{tabular}{|c|c|c|c|c|c|}\hline
Image ${\cal I}$&Number of points of $I$&Time for computing &$\beta_0$&$\beta_1$&$\beta_2$\\
\hline
$I_1$&$26308$&  $50$ seconds &$2$&$9$&$3$\\
$I_2$&$31012$& $38$ seconds &$138$&$419$&$13$
\\
$I_3$&$18842$&$27$ seconds&$1$&$277$&$5$\\
\hline
\end{tabular}
\end{center}

\begin{figure}[t!]
\centerline{\includegraphics[width=12cm]{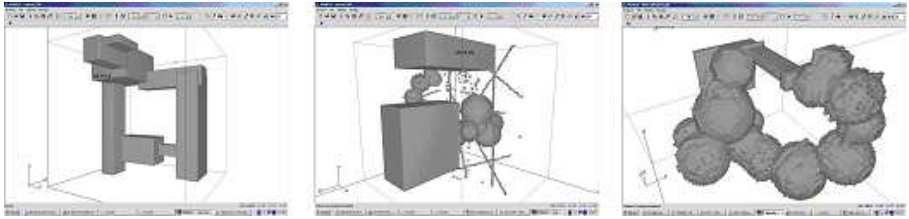}}
\caption{The 3D digital images $I_1$, $I_2$ and $I_3$.}
\end{figure}

\subsection{Computing ``Good" Representative Cycles of Homology Generators}

In \cite{EW05}, algorithms for obtaining
``optimal" generators of the first homology group are developed
using Dijkstra's shortest path algorithm for any oriented
$2$-manifolds.
The algorithms presented in \cite{gbrsamuel}  produce ``nice" representative cycles of homology generators that
always fit on the boundaries of the image.

We sketch here some techniques for drawing ``good"
representative cycles of homology generators in the context of
digital volumes.
Let $(C_{\scst I},\phi_{\scst I})$ be an AM-model for a 3D digital image ${\cal I}=({\cal V},I,14,14)$.
We denote by $\partial{\cal I}=({\cal V},\partial I,14,14)$ the digital image
such that  $\partial I$ is the set of points of $I$ with at least one $14$-neighbor white point of ${\cal I}$.
We say that $M$ is a set of ``good" representative cycles of
homology generators of $I$ if each chain  of $M$ satisfies that:
\begin{itemize}
\item it is a cycle;
\item it belongs to ${\cal C}(K(\partial I))$;
\item in
dimension $0$, it is a vertex;
\item in dimension $1$, it is an {\em
elementary cycle} (that is, it is connected, each vertex is shared by
exactly two edges and two consecutive edges can not belong to the
same triangle in $K(I)$);
\item  in dimension $2$, it is an {\em
elementary  cavity} (that is, it is a connected   $2$-cycle with exactly
one white connected component inside and three triangles can not
belong to the same tetrahedra in $K(I)$).
\end{itemize}
First of all, suppose we have an AM-model for $\partial {\cal I}$, $(C_{\scst \partial I}, \phi_{\scst \partial I})$,
and ${\cal I}$, $(C_{\scst I}, \phi_{\scst I})$.
Let
$\pi_{\scst \partial I}=id-\partial\phi_{\scst \partial I}-\phi_{\scst \partial I}\partial$ and
$\pi_{\scst I}=id-\partial\phi_{\scst I}-\phi_{\scst I}\partial$.
Let us denote by $\{\alpha_1,\dots,\alpha_n\}$
 the elements of $M_{\scst \partial I}$ which is a base of
 ${\cal M}_{\scst \partial I}=$ im $\pi_{\scst \partial I}$. Suppose that
 if $h\in M_{\scst \partial I}$ then $\partial(h)=0$ (we can obtain this if we compute the AM-model for $\partial I$
 using Algorithm \ref{ammodel}).
 The cycles of $M_{\scst \partial I}$ are representative cycles of the homology generators of $\partial I$.
Decompose and replace each $0$-cycle in $M_{\scst \partial I}$ by its constitutive vertices, each
$1$-cycle in $M_{\scst \partial I}$ by its elementary  cycles  and each
$2$-cycle in $M_{\scst \partial I}$ by its elementary cavities.
Note that all the  homology generators  of $I$
are homology generators of $\partial I$,
therefore a set of representative cycles of the homology generators of $I$ is a subset of $M_{\partial I}$.
Let $M'_{\scst I}$ be
the set $\{\alpha: \; \alpha\in  M_{\scst \partial I}$ and $\pi_{\scst I}(\alpha)=\alpha\}$.
Obtain the new AM-model $(C_{\scst I}, \phi'_{\scst I})$ applying Lemma \ref{lema1} to the elements of the set
$M'_{\scst I}$.
Then, $(C_{\scst I}, \phi'_{\scst I})$
is a new AM-model for ${\cal I}$; and $M'_{\scst I}$ is a base of ${\cal M}_{\scst I}=$ im $\pi_{\scst I}$ and
a set of good representative cycles of  homology generators  of $I$.

\begin{figure}[t!]
\centerline{\includegraphics[width=12cm]{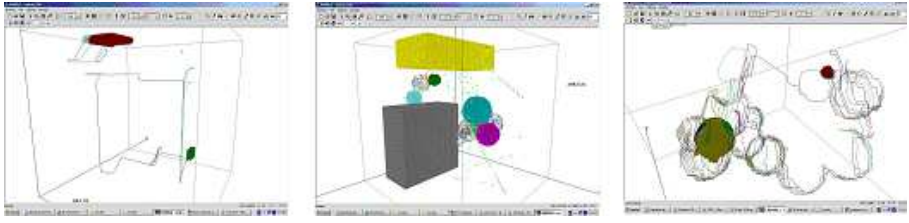}}
\caption{Representative cycles of the homology generators of the images $I_1$, $I_2$ and $I_3$.}
\end{figure}

\subsection{AM-models after Adding or Deleting a Voxel}

Now, we study the problem of
 topologically
controlling  a digital image using  AM-models  when it suffers local changes.
More concretely, we show how to compute  an AM-model
for a digital image  when a voxel is added or deleted using the AM-model computed before.
Adding or deleting a voxel $v$ of ${\cal I}$ means to change the color of a point $v$ in ${\cal I}$
and it consists of
 adding or deleting  a set of simplices  having $v$ as a vertex.
Since we work with simplicial complexes representing 3D digital images considering the 14-adjacency,
the maximum number of simplices
having $v$ as a vertex is $74$.
The key idea of both algorithms is that when a $q$-simplex $\sigma$ is added to or deleted from an
AM-model for a simplicial complex $K$,  we only have to put into the SNF the column of the  matrix
of $\partial_q$ relative to $\sigma$ to obtain the new AM-model.

\noindent{\bf AM-models after Adding a  Voxel}.
Given a digital image ${\cal I}=({\cal V},I,14,14)$, we add to $I$ a point $v\not\in I$ to obtain a new
digital image denoted by ${\cal I}^{\cup v}=({\cal V}, I\cup\{v\},14,14)$. Therefore,
the  addition of a point $v$ to $I$  consists in the addition to $K(I)$
of  all the simplices of   $K(I\cup\{v\})$ having $v$ as a vertex. In each step of the process,
 one simplex is added.
Given an AM-model for a digital image ${\cal I}=({\cal V},I,14,14)$ such that $I$ has $m$
points, the following algorithm computes an AM-model for ${\cal I}^{\cup v}$, with integer coefficients, in ${\cal O}(m^2)$

\begin{alg}\label{incremental}
An Incremental Algorithm for Computing an AM-model for a 3D Digital Image ${\cal I}$.
\begin{tabbing}
{\sc Input:} \= {\tt A digital image $({\cal V},I,14,14)$,
a point $v\not\in I$ and an AM-model}\\
\> {\tt  $(C_{\scst I}, \phi_{\scst I})$ for ${\cal I}$
such that,  in each dimension $q$,
the}\\
\> {\tt  matrix of $\partial_q$ with
 respect to $C_{\scst I}$ coincides with its SNF.
}\\
 {\tt Let $\{\sigma_1,\dots,\sigma_n\}$ ($n\leq 74$) be
the  ordered-by-increasing-dimension }\\
{\tt  set of all the simplices of $K(I\cup\{v\})$ having $v$ as a vertex.}\\
{\tt $C^{\cup v}:=C_{\scst I}$ and $\phi^{\cup v}:=\phi_{\scst I}$.}\\
{\tt For }\= {\tt $i=1$ to $i=n$ do:}\\
\>{\tt Let $q$ be the dimension of $\sigma_i$, let $C^{\cup v}_q=\{a_1,\dots,a_r\}$ and}\\
\>{\tt $C^{\cup v}_{q-1}=\{e_1,\dots,e_s\}$ such that  $\partial_q(a_j)=e_j$ for $1\leq j\leq t$, }\\
\>{\tt  $\partial_q(a_j)=0$  for
$t< j\leq r$ and    $\partial_q(\sigma_i)=\sum_{\ell=1}^{s}\rho_{\ell}e_{\ell}$ for some $\rho_{\ell}\in {\bf Z}$.}\\
 \>{\tt $a:=\sigma_i-\sum_{\ell=1}^{t}\rho_{\ell}a_{\ell}$ and  $C^{\cup v}_q:=\{a_1,\dots,a_r,a\}$.}\\
\>{\tt If }\= {\tt $\rho_{\ell}=0$ for $\ell > t $ then $\phi^{\cup v}(a):=0$.}\\
\>{\tt Else }\= {\tt  obtain the SNF of
the matrix of  $\partial_q$}\\
\>\>{\tt  relative to    some base $C^{\cup v}_{q-1}:=\{e_1,\dots,e_t,e'_{t+1},\dots,e'_{s}\}$ then}\\
\>\>{\tt
  $\phi^{\cup v}(a):=0$, $\phi^{\cup v}(e'_{t+1}):=a$ and  $\phi^{\cup v}(e'_{j}):=0$ for the rest.}\\
{\sc Output:} \= {\tt An AM-model $(C^{\cup v},
\phi^{\cup v})$ for ${\cal I}^{\cup v}$. }\\
\end{tabbing}
\end{alg}

\noindent{\bf AM-models after Deleting a Voxel}.
Given a digital image ${\cal I}=({\cal V},I,14,14)$, we delete from $I$ a point $v\in I$ to obtain a new
digital image denoted by ${\cal I}^{\setminus v}=({\cal V}, I\setminus\{v\},14,14)$. Therefore,
the  deletion of a point $v$ from $I$  consists in the deletion from $K(I)$
of  all the simplices of   $K(I)$ having $v$ as a vertex. In each step of the process,
 one simplex is deleted.
Given an AM-model for a digital image ${\cal I}=({\cal V},I,14,14)$ such that $I$ has $m$
points, the following algorithm computes an AM-model for ${\cal I}^{\setminus v}$, with integer coefficients, in ${\cal O}(m^2)$

\begin{alg}\label{decremental} A Decremental Algorithm for Computing an AM-model for a 3D Digital Image ${\cal I}$.
\begin{tabbing}
{\sc Input:} \= {\tt A digital image $({\cal V},I,14,14)$,
a point $v\in I$ and an AM-model}\\
\> {\tt  $(C_{\scst I}, \phi_{\scst I})$ for ${\cal I}$
such that  in each dimension $q$,}\\
\> {\tt  the matrix of $\partial_q$ with
 respect to $C_{\scst I}$ coincides with its SNF.
}\\
{\tt Let $\{\sigma_1,\dots,\sigma_n\}$ ($n\leq 74$) be the ordered-by-decreasing-dimension }\\
{\tt  set of all the simplices of $K(I)$ having $v$ as a vertex.}\\
{\tt $K:=K(I)$, $C^{\setminus v}:=C_{\scst I}$ and $\phi^{\setminus v}:=\phi_{\scst I}$.}\\
{\tt For }\= {\tt $i=1$ to $i=n$ do}\\
\>{\tt Let $q$ be the dimension of $\sigma_i$, let $C^{\setminus v}_q=\{a_1,\dots,a_r\}$ and}\\
\>{\tt $C^{\setminus v}_{q-1}=\{e_1,\dots,e_s\}$ such that  $\partial_q(a_j)=e_j$ for $1\leq j\leq t$,  $\partial_q(a_j)=0$}\\
\>{\tt for $t< j\leq r$ and $\partial_q(\sigma_i)=\sum_{\ell=1}^{s}\rho_{\ell}e_{\ell}$ for some
$\rho_{\ell}\in {\bf Z}$.}\\
\>{\tt Let }\= {\tt $k$ be  the smallest index  such that $\{a_1,\dots,\hat{a_k},\dots,a_r\}$ is}\\
\>{\tt  a base of ${\cal C}_q(K\setminus\{\sigma_i\})$ then}\\
\>\> {\tt $C^{\setminus v}_{q}:= \{a_1,\dots,\hat{a_k},\dots,a_r\}$  and $\phi(e_k):=0$.}\\
\>{\tt $K:=K\setminus \{\sigma_i\}$.}\\
{\sc Output:} \= {\tt An AM-model $(C^{\setminus v},\phi^{\setminus v})$ for ${\cal I}^{\setminus v}$. }
\end{tabbing}
\end{alg}

Observe that if an AM-model $(C, \phi)$ for ${\cal I}$ is computed using Algorithm \ref{ammodel},
\ref{incremental} or \ref{decremental}
then
in each dimension $q$, it satisfies that the  matrix of $\partial_q$ with
 respect to $C$ coincides with its SNF.

\subsection{AM-models for  3D Digital Images under Voxel-Set Operations}

In this subsection, we reuse the AM-model information for digital images under
voxel-set operations (union, intersection, difference and inverse).
Let ${\cal I}=({\cal V},I,14,14)$ and ${\cal J}=({\cal V},J,14,14)$ be two digital images, then
${\cal I}\cup {\cal J}=({\cal V},I\cup J,14,14)$,
${\cal I}\cap {\cal J}=({\cal V},I\cap J,14,14)$,
${\cal I}\setminus {\cal J}=({\cal V},I\setminus J,14,14)$.
In order to define the inverse of ${\cal I}$,
for each $p=(x_1,x_2,x_3)\in {\cal V}$, let $X_{p}=$ max$\{|x_i|:\; i=1,2,3\}$.
Let $X_{\scst I}=$ max$\{ X_p:\; p\in I\}$.
Let  ${\cal G}_{\scst I}$ be the digital image $({\cal V},G_{\scst I},14,14)$
where $G_{\scst I}=\{p:\; p \in {\cal V}$ and $X_p\leq X_{\scst I}+1\}$. Then, the inverse of ${\cal I}$,
$\bar{\cal I}$, is ${\cal G}_{\scst I}\setminus {\cal I}$.
 We will not consider any of the trivial cases
 $I=\emptyset$, $J=\emptyset$, $I\cap J= \emptyset$, $I\subseteq J$,
  or $I=\{p:\; p \in {\cal V}$ and $X_p\leq r\}$ for some $r\in {\bf Z}$.

Let ${\cal L}=({\cal V},L,14,14)$ be a digital image and $F=\{v_1,\dots,v_m\}\subset{\cal V}$ such that
$F\subset L$ or $F\cap L=\emptyset$. If $F\subset L$, denote by ${\cal L}^{F}$ the
image $({\cal V},L\setminus F,14,14)$. On the other hand, if $F\cap L=\emptyset$, denote by ${\cal L}^{F}$
the image $({\cal V},L\cup F,14,14)$. Let $(C_{\scst L},\phi_{\scst L})$ be an AM-model for ${\cal L}$
such that in each dimension $q$,
the  matrix of $\partial_q$ with
 respect to $C_{\scst L}$ coincides with its SNF.
 Algorithm \ref{pre} is a common processing  to the four voxel-set operations treated here.

\begin{alg}\label{pre} Common Processing.
\begin{tabbing}
{\sc Input:} \= {\tt  The AM-model $(C_{\scst L},\phi_{\scst L})$ for
${\cal L}=({\cal V},L,14,14)$
and the set of} \\
\> {\tt points  $F=\{v_1,\dots,v_m\}$ such that $F\subset L$ or $F\cap L=\emptyset$}.\\
{\tt If }\= {\tt $F\subset L$ then}\\
\> {\tt For }\= {\tt $i=1$ to $i=m$ do}\\
 \>\>{\tt apply Algorithm \ref{decremental} to $v_i$ and the AM-model
$(C_{\scst L},\phi_{\scst L})$.}\\
\>\> {\tt $C_{\scst L}:=C_{\scst L}^{\setminus v_i}$ and $\phi_{\scst L}:=\phi_{\scst L}^{\setminus v_i}$.}\\
{\tt Else }\=  {\tt for }\= {\tt $i=1$ to $i=m$ do}\\
 \>\>{\tt apply Algorithm \ref{incremental} to $v_i$ and the AM-model
$(C_{\scst L},\phi_{\scst L})$.}\\
\>\> {\tt $C_{\scst L}:=C_{\scst L}^{\cup v_i}$ and $\phi_{\scst L}:=\phi_{\scst L}^{\cup v_i}$.}\\
{\tt $C_{\scst L}^{\scst F}:=C_{\scst L}$ and $\phi_{\scst L}^{\scst F}:=\phi_{\scst L}$}\\
{\sc Output: }\= {\tt
 An AM-model $(C_{\scst L}^{\scst F},\phi_{\scst L}^{\scst F})$ for  ${\cal L}^{F}$.}
\end{tabbing}
\end{alg}

Let
$(C_{\scst I}, \phi_{\scst I})$ and
$(C_{\scst J}, \phi_{\scst J})$ be an AM-model
for ${\cal I}$ and ${\cal J}$, respectively, such that  in each dimension $q$,
the  matrix of $\partial_q$ with
 respect to $C_{\scst L}$, for $L=I,J$, coincides with its SNF.
The following algorithm computes an AM-model for ${\cal I}\cup {\cal J}$.

\begin{alg} Computing an AM-model for ${\cal I}\cup {\cal J}$.
\begin{tabbing}
{\sc Input:} \= {\tt The AM-models $(C_{\scst I}, \phi_{\scst I})$ for ${\cal I}$ and
$(C_{\scst J},\phi_{\scst J})$ for ${\cal J}$. }\\
{\tt Apply Algorithm \ref{pre} to $(C_{\scst L}, \phi_{\scst L})$ and $F:=I\cap J$ for $L=I,J$.}\\
{\tt For }\= {\tt each $a\in C:=C_{\scst I}^{\scst F}\cup C_{\scst J}^{\scst F}$}\\
 \>{\tt  $\phi(a):=\phi^{\scst F}_{\scst I}(a)$ if $a\in C_{\scst I}^{\scst F}$;
and  $\phi(a):=\phi_{\scst J}^{\scst F}(a)$ if $a \in C_{\scst J}^{\scst F}$. }\\
{\tt Apply Algorithm \ref{pre} to $F$ and the
AM-model $(C,\phi)$.}\\
{\tt $C_{\scst I\cup J}:=C^{\scst F}$ and $\phi_{\scst I\cup J}:=\phi^{\scst F}$. }\\
{\sc Output: }\= {\tt An AM-model $(C_{\scst I\cup J},\phi_{\scst I\cup J})$ for ${\cal I}\cup {\cal J}$.}
\end{tabbing}
\end{alg}

Algorithm \ref{pre} is the essential step for computing an AM-model for ${\cal I}\cap {\cal J}$, ${\cal I}\setminus {\cal J}$ and ${\cal I}$.

\begin{alg}\label{cap} Computing an  AM-model for  ${\cal I}\cap {\cal J}$.
\begin{tabbing}
{\sc Input:} \= {\tt The  AM-model $(C_{\scst I}, \phi_{\scst I})$ for ${\cal I}$ and the set
$F=I\setminus J$.} \\
{\tt Apply Algorithm \ref{pre} to   $(C_{\scst I}, \phi_{\scst I})$ and
$F$.}\\
  {\sc Output: }\= {\tt An AM-model $(C_{\scst I}^{\scst F},\phi_{\scst I}^{\scst F})$ for  ${\cal I}\cap \bar{\cal J}$.}
\end{tabbing}
\end{alg}

An algorithm for computing an AM-model for ${\cal I}\setminus {\cal J}$ (resp. for $\bar{\cal I}$) is similar to the one above.
The only difference is that the input is
an AM-model $(C_{\scst I}, \phi_{\scst I})$ for ${\cal I}$ and the set
$F=I\cap J$ (resp. an AM-model $(C_{G_I}, \phi_{G_I})$ for ${\cal G}_I$ and the set $I$).

\section{Comments and Future Work}

The (non-unique) algebraic-topological representation of
a given simplicial complex of any dimension showed here, allows  to
compute integer (co)homology, representative cycles of integer (co)homology generators, the cup product on cohomology
with integer coefficients
 and
a topological invariant derived from the integer
cohomology ring. Moreover, we give a positive answer to the
problem of  reusing AM-models for determining
homological information of new 3D binary digital images
constructed from the previous ones using voxel-set operations.

There is considerable scope for further research:
\begin{itemize}
\item To extend our
method to nD binary digital images in any grid using simplicial
analogous techniques \cite{KRR92,KKM90,ADFQ00}.
\item To compute topological invariants from primary cohomology operations in the discrete
setting of digital images. The work done in \cite{GR01} seems to be a compulsory reading for advancing in
this issue.
\item To compute homotopy groups of nD binary digital images or $n$-$G$-maps \cite{Lachaud} using AM-models. The works
 \cite{Mal01,KRR92,Rea00,F87,ADFQ00} could help us in
this task.
\item To develop a discrete Morse theory \cite{For95}
for digital images well-adapted to our method. The paper
\cite{Sko06} would be  a good starting point.
\end{itemize}

Potential applications of our  method in Computer Vision and Digital Image Processing involving
 not only
3D object but also higher dimensional structures can be encountered in Medical Imaging and Object Modelling.
Moreover, our method seems to be especially well-adapted to segmentation under topological constraints
 and elimination of small topological noise.

\end{document}